\documentclass[conference, a4paper]{IEEEtran}
\IEEEoverridecommandlockouts
\usepackage{cite}
\usepackage{amsmath,amssymb,amsfonts}
\usepackage{algorithmic}
\usepackage{algorithm}
\newcommand{\myCOMMENT}[1]{\textcolor{darkgreen}{\texttt{// #1}}}

\usepackage{subcaption}
\usepackage{placeins}
\usepackage[labelfont=bf]{caption}

\usepackage{graphicx}
\usepackage{textcomp}
\usepackage{xcolor}
\usepackage{color}
\usepackage{hyperref}

\hypersetup{
    colorlinks=true,
    linkcolor=blue,
    filecolor=magenta,      
    urlcolor=cyan,
    citecolor = magenta,
    pdfpagemode=FullScreen,
}

\definecolor{darkgreen}{rgb}{0,0.5,0}

\def\BibTeX{{\rm B\kern-.05em{\sc i\kern-.025em b}\kern-.08em
    T\kern-.1667em\lower.7ex\hbox{E}\kern-.125emX}}

\usepackage{printlen}
\uselengthunit{mm}

\usepackage{layouts}
    
\begin{document}





\title{SATA: \underline{S}parsity-\underline{A}ware Scheduling for Selective \underline{T}oken \underline{A}ttention}




\author{
\IEEEauthorblockN{ 
\centering
Zhenkun Fan\textsuperscript{1},
Zishen Wan\textsuperscript{1},
Che-Kai Liu\textsuperscript{1},
Ashwin Sanjay Lele\textsuperscript{2},
Win-San Khwa,\textsuperscript{3},
Bo Zhang\textsuperscript{2}}
\IEEEauthorblockN{
Meng-Fan Chang\textsuperscript{3},
Arijit Raychowdhury\textsuperscript{1}
}
\IEEEauthorblockA{
\textsuperscript{1}\textit{Electrical and Computer Engineering Department, Georgia Institute of Technology}, Atlanta, GA, USA \\
\textsuperscript{2}\textit{TSMC Corporate Research}, San Jose, CA, USA \\
\textsuperscript{3}\textit{TSMC Corporate Research}, Hsinchu, Taiwan \\
}

{
Email: \{zfan87, zwan63\}@gatech.edu
}
}


\maketitle



\begin{abstract}

Transformers have become the foundation of numerous state-of-the-art AI models across diverse domains, thanks to their powerful attention mechanism for modeling long-range dependencies. However, the quadratic scaling complexity of attention poses significant challenges for efficient hardware implementation. While techniques such as quantization and pruning help mitigate this issue, \emph{selective token attention} offers a promising alternative by narrowing the attention scope to only the most relevant tokens, reducing computation and filtering out noise.

In this work, we propose SATA, a \emph{locality-centric dynamic scheduling scheme} that proactively manages sparsely distributed access patterns from selective Query-Key operations. By reordering operand flow and exploiting data locality, our approach enables early fetch and retirement of intermediate Query/Key vectors, improving system utilization. We implement and evaluate our token management strategy in a control and compute system, using runtime traces from selective-attention-based models. Experimental results show that our method improves system throughput by up to 1.76$\times$ and boosts energy efficiency by 2.94$\times$, while incurring minimal scheduling overhead.

\makeatletter
\typeout{PAPERWIDTH=\the\paperwidth}
\typeout{PAPERHEIGHT=\the\paperheight}
\typeout{TEXTWIDTH=\the\textwidth}
\typeout{TEXTHEIGHT=\the\textheight}
\typeout{COLUMNWIDTH=\the\columnwidth}
\typeout{COLUMNSEP=\the\columnsep}
\typeout{BASEFONTSIZE=\f@size pt}
\makeatother

\end{abstract}


\section{Introduction}
\label{intro}
Since Transformers revolutionized natural language processing (NLP) with their breakout success across multiple tasks~\cite{vaswani2017attention}, the attention mechanism has become foundational in modern machine learning. Transformers now power state-of-the-art models in computer vision~\cite{dosovitskiy2020image,liu2021swin,touvron2021training,wan2025generative}, natural language processing~\cite{devlin2018bert,wei2022chain,radford2018improving,xie2025realm}, multimodal learning and agentic tasks~\cite{team2023gemini,lu2019vilbert,seo2022end,wan2025reca,wang2025slm}, laying the groundwork for the next generation of AI algorithms.
Despite their impressive capabilities, Transformers come with high deployment overhead, often requiring multiple GPUs and incurring substantial energy consumption~\cite{schwartz2020green}.


The power of attention lies in its global receptive field, allowing each token to attend to all others within a context. However, this expressiveness also leads to the well-known quadratic complexity in sequence length, making the attention layer a computational bottleneck. As model sizes grow, attention increasingly becomes memory-bound~\cite{kim2023full,ibrahim2024special}, posing challenges for scaling to edge platforms. Consequently, both academia and industry have been actively developing Transformer acceleration techniques to improve throughput and energy efficiency~\cite{fan2022adaptable,ham2021elsa,dao2023flashattention,roy2025breaking}.


To reduce redundant computation, recent work has explored sparsity in Transformers. These models often exhibit inherent redundancy in token, head, or weight-level representations. Sparse Transformer accelerators~\cite{ham20203, wang2021spatten, zhou2022energon, tuli2023acceltran} have leveraged this property to reduce workload and boost efficiency while maintaining accuracy. In particular, TopK Selective Query-Key Attention~\cite{zhao2019explicit, wang2022kvt, chen2023learning, singh2024tosa, xiao2024ttst} has gained traction by focusing on the most influential key tokens for a given query, thereby pruning less important MAC operations in the attention score computation. This algorithmic sparsity opens new opportunities for hardware-software co-design to optimize performance.


At the core of the attention mechanism lies the matrix multiplication (MatMul) primitive. As a fundamental building block of ML workloads, MatMul has seen significant hardware optimization over the past decade. Compute-in-Memory (CIM) architectures are a notable direction, offering latency and energy benefits by minimizing memory movement and enabling massive parallelism. Recent Transformer accelerators have extended CIM capabilities with features like interconnect reconfigurability~\cite{fu2023p, tu2022trancim}, bitwidth adaptability~\cite{tu2023multcim}, and algorithm-hardware co-design~\cite{yang2020retransformer}. 
Due to its dynamic nature, selective Q-K attention often leads to fragmented operational flow. Harvesting energy savings by halting the corresponding functional unit brings down hardware utilization. 
In the worst case, the reduced operand reuse distance can induce a surge of external memory access, degrading system efficiency and throughput.  


In this work, we propose SATA, a \emph{locality-centric dynamic scheduling scheme} that reorders Query and Key access patterns to improve operand locality in attention computation. 
Our scheduler targets Scaled Dot-Product Attention (SDPA) across Multi-Head Attention (MHA) layers, executing only the most impactful QK-MAC operations while maximizing system utilization. 
Through centralizing Multiply and Accumulates~(MACs), we minimized the retention duration of Q/K operand without sacrificing model accuracy.
Combined with an optimized scheduler architecture, the system utilization is maximized with parallelized Q, K operations


This paper, therefore, makes the following contributions:

\begin{itemize}
    \item We propose a sparsity-aware scheduling scheme that enhances data reuse in sparse Query-Key attention. By minimizing reuse distance through sorted operand access, the scheme improves throughput and energy efficiency.
    \item We design a lightweight controller to implement the proposed scheduling scheme within digital design flow. Experimental results show the scheduling overhead is just 2.2\% in the most energy-sensitive workload, with a worst-case overhead of 5.9\%.

    \item We develop a comprehensive benchmarking framework using a silicon-validated CIM simulator and metadata from commercial EDA tools, evaluating across four Transformer models. Our design achieves throughput gains of 1.47$\times$, 1.76$\times$, 1.59$\times$, and 1.5$\times$, and energy efficiency improvements of 1.81$\times$, 2.1$\times$, 1.85$\times$, and 2.94$\times$ on TTST, KVT-DeiT-Tiny, KVT-DeiT-Base, and DRSformer, respectively. The code is publicly available at~\href{https://github.com/SenFFF/SATA}{github.com/SenFFF/SATA}.
\end{itemize}


\section{Background and Related Work}
\label{sec:background}

\begin{figure}
\centering
    \includegraphics[width=.9\linewidth]{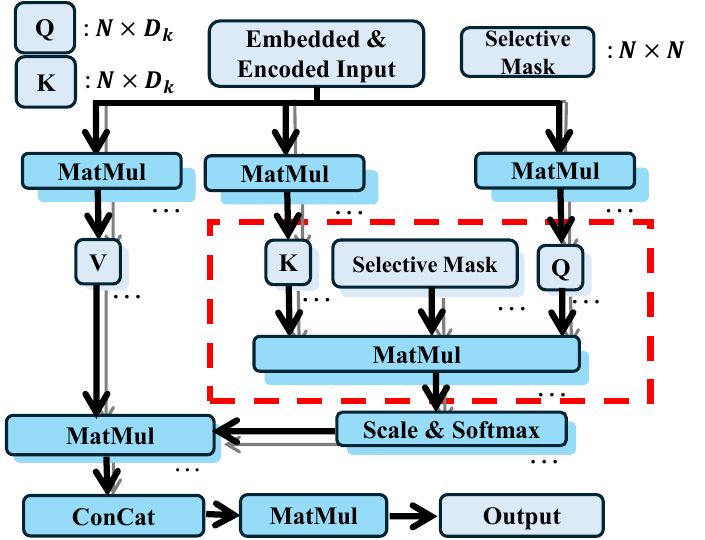}
    \caption{\textbf{MHA visualization.} Red dashed box represents the targeted workload in SATA. $N$: sequence length (number of tokens); $D_k$: embedding dimension of Query and Key.}
    \vspace{-1pt}
    \label{fig:MHA}
\end{figure}

\subsection{Attention Mechanism}
The attention mechanism (Fig.~\ref{fig:MHA}) is first introduced in Neural Machine Translation~\cite{bahdanau2014neural}.
In a series of tokenized inputs, each token searches through the sequence and generates scores for other tokens. It then forms a feature representation with MAC upon vectors embedded in Value space weighted by the relevance scores.
Despite state-of-the-art performance, modern transformers grow exponentially in terms of size and cost.
Commercial products rely heavily on datacenters. Transmission between cloud and edge has incurred great overhead and security concerns.
Synergistic approaches to slim down transformers are a heated topic in academia and industry.

\subsection{Attention Sparsity}
Transformer tokens are projected to high-dimensional spaces before further proceeding. 
Such redundancy is vital to training.
Inference-centric models, on the other hand, have an effective datapath where tokens can be identified across the spectrum from critical to obsolete. 
Identifying inessential components and avoiding related compute have great potential for transformer acceleration.

Among all techniques, TopK selective attention is a more flexible scheme where each Query(Q) is computed against only a subset of Keys(K)~\cite{zhao2019explicit, wang2022kvt}. 
As dense transformers are often considered overparameterized, selective attention serves as a regularization method providing robust performance and sheds light on a path for efficient hardware implementation.

\subsection{Transformer Accelerators}
Accelerating transformers has been one of the hottest topics in the past decade. 
For sparsity-driven approaches, A$^3$~\cite{ham20203} performed successive approximation.
SpAtten~\cite{wang2021spatten} combined cascaded token pruning, head pruning, TopK and dynamic quantization to alleviate DRAM access.
Energon~\cite{zhou2022energon} implemented selective attention by progressively filtering trivial keys.
Inspired by the dynamic philosophy in~\cite{tu2023multcim}, SATA performed dynamic scheduling in MHA (Fig.~\ref{fig:MHA}) to improve throughput and system efficiency.


\section{Scheduling Scheme}
\label{sec:alg}

\begin{figure*}
    \centering
    \includegraphics[width=0.85\linewidth]{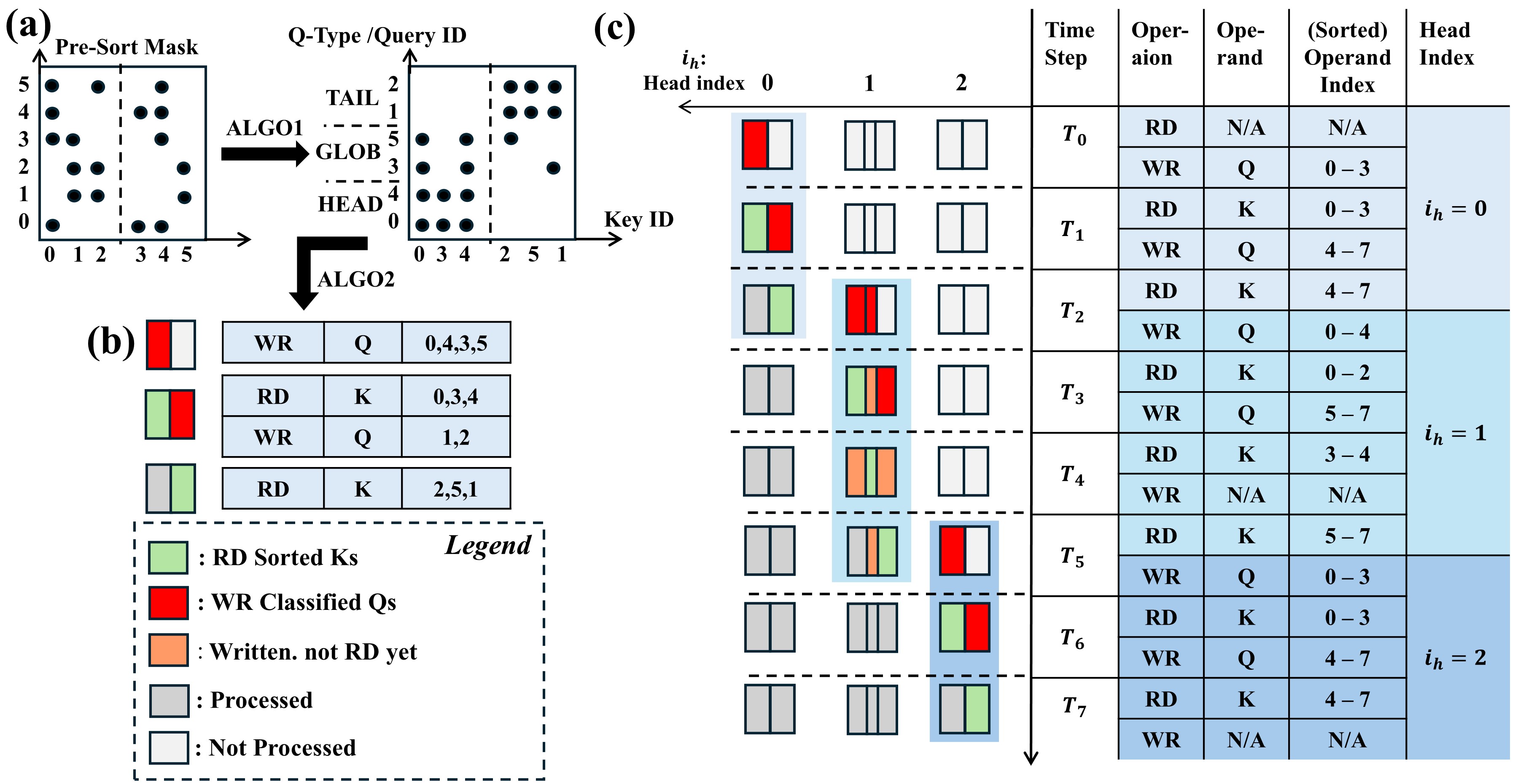}

    \caption{\textbf{Visualization for SATA algorithm.} (a),(b) present a demo for Algo.~\ref{alg:sort},~\ref{alg:schedule} ($N=6$, $S_h=\frac{N}{2}$). Q/K indices are original token indices. This head is classified as \underline{HEAD} since the number of \underline{GLOB} Qs is smaller than $\frac{N}{2}$. Tie is broken by assigning head condition to \underline{HEAD} when number of HEAD-Qs equals that of TAIL-Qs. (c) presents a scheduled demo. Two heads ($i_h=0,2$) are perfectly sorted with $S_h=\frac{N}{2}$. The rest ($i_h=1$) represents a more general case ($S_h<\frac{N}{2}$) where conceding ($S_h$ decrement) has happened.
    }
    
    \label{fig:schedule_temporal}
\end{figure*}

In this section, we present the SATA scheduling scheme. We begin by characterizing workload, then describe the algorithmic flow, and finally cover the hardware implementation. The notation for parameters can be found in Algo.~\ref{alg:sort},~\ref{alg:schedule} and Tab.~\ref{tab:model_table}.

\subsection{Selective Attention Layer}
MHA can be decomposed to projection, Q-K MatMul, A-V MatMul, Feed Forward Network (FFN), and nonlinear operations. 
Projection and FFN are considered Static MatMul. 
In contrast, Q-K and A-V MatMuls rely on vectors generated on the fly and are input-dependent. These dynamic MatMul scale non-linearly, hindering attention's edge deployment~\cite{kim2023full}.

We focus therefore on Q-K MatMul across heads. 
Here, the input to SATA is the TopK indices of Keys relevant to Queries. 
The acquisition of indices has been studied extensively~\cite{wang2021spatten,ham2021elsa,zhou2022energon} and its cost is integrated in evaluation (Sec.~\ref{sec:eva}).


\begin{algorithm}
{\footnotesize
    \caption{Key Sorting and Query Classifying}
    \label{alg:sort}
    \begin{algorithmic}[1]
        \STATE \textbf{Input:} \#Token=$N$, Selective Mask $QK \in \{0,1\}^{N \times N}$, threshold $\theta$
        \STATE \textbf{Output:} Sorted Key-Order \texttt{Kid} , Q-Types $QT$, Head-Type $HT$.
        \STATE
        
        \STATE \texttt{Kid} = [], $Dummy = \mathbf{0}^{N}$ 
        \myCOMMENT{Init}
        \STATE \myCOMMENT{Rand Seed}
        \STATE $\_kid$ = rand($N$), $Dummy$.\text{update}$(QK[:,\_kid])$, $\texttt{Kid}.\text{append}(\_kid)$  

        \STATE \myCOMMENT{Sort}
        \FOR{$i$ in $0$ ... $N-2$}  
            \STATE $\_kid \gets \arg\max\limits_i (Dummy^T \cdot QK[:,i]), i\notin Kid$

            \STATE $\texttt{Kid}.\text{append}(\_kid)$
            \STATE $Dummy.\text{update}(QK[:,\_kid])$    
        \ENDFOR

        \STATE \myCOMMENT{Classify}
        \STATE $S_h = \frac{N}{2}$, $QK_s = QK[:,\texttt{Kid}]$, $QT = []$, \#Glob = 0
        \WHILE{True}
            \FOR{$i$ in $0$ ... $N-1$}
                \STATE $QT.\text{update}(\text{classify}(QK_s[i,:], S_h ))$
                \STATE \#Glob += 1 \textbf{if} (i-th Q is \underline{GLOB})
            \ENDFOR

            \IF{\#Glob $>$ $\theta$}
                \STATE $S_h \gets S_h-1$ \myCOMMENT{Escape \underline{GLOB} w/ smaller $S_h$}
                \STATE \textbf{continue}
            \ELSE
                \STATE $HT \gets \#(\text{HEAD}) > \#(\text{TAIL}) ? \text{\underline{HEAD}} : \text{\underline{TAIL}}$
                \STATE \textbf{Break}
            \ENDIF
        \ENDWHILE

        \STATE \textbf{return} \texttt{Kid}, $QT$, $S_h$, $HT$
    \end{algorithmic}
}
\end{algorithm}

\subsection{Intra-Head Mask Sorting}


Fig.~\ref{fig:schedule_temporal}(a) shows how SATA sorts selective masks.
Scattered K access often leads to redundant operand fetches.
SATA introduces intra-head mask sorting, where Qs and Ks are reordered to improve operand locality before scheduling. 

The process begins by randomly picking a K index as the pointer. 
Its access pattern~(mask column), copied as vector \textit{dummy}, serves as a reference vector throughout sorting. 
Unsorted K access vectors are compared with \textit{dummy} to generate similarity scores.
The K index with the highest score would be marked as the next K pointer, and its access pattern is used to increment \textit{dummy}. 
This process loops till all Keys are sorted and generates a list of sorted K indices. (Algo.~\ref{alg:sort}, line 4-12)
Despite the order of $O(n^2)$, the overhead of sorting is light as the mask is binary. 
Sorting is further optimized in Sec.~\ref{sec:hw_describe}.

With sorted Ks, Qs would be classified into three groups: HEAD, TAIL or GLOB, based on sorted distribution.
Classification relies on a dynamic 'Heavy Size' $S_h$. Specifically,
\begin{itemize}
    \item Qs not accessing last $S_h$ sorted Ks are tagged as HEAD.
    \item Qs not accessing first $S_h$ sorted Ks are tagged as TAIL.
    \item Other Qs are tagged GLOB which represents less locality.
\end{itemize}

A head is local as long as GLOB Qs do not dominate. 
The exact type depends on the dominant Q-type being HEAD or TAIL (head-type \underline{HEAD} or \underline{TAIL}). 
Otherwise, the head is in a \underline{GLOB} state. 
Should that happen, $S_h$ is decremented and Qs are reclassified to escape from GLOB status (Algo.~\ref{alg:sort}, line 14-27).
Packed together, this two-step process enhances data reuse and lays the foundation for sparsity aware scheduling.



\subsection{Sparsity-Aware Inter-Head Scheduling}
\label{subsec:schedule}

\begin{algorithm}
{
\footnotesize
    \caption{Sparsity-Aware Scheduling}
    \label{alg:schedule}
    \begin{algorithmic}[1]
        \STATE \textbf{Input:} \#Token=$N$, \#Heads=$N_h$, head-Types $HT$, \newline sorted Key indices for all heads \texttt{Kid}, $\in \mathbb{N}^{N_h \times N}$, classified Q-Types $QT_i=[Q_{i,\text{HEAD}}; Q_{i,\text{TAIL}}; Q_{i,\text{GLOB}}], i\in[0, \ldots ,(N_h-1)];$, 
        Heavy-Size $S_{h,i}, i\in[0, \ldots ,(N_h-1)];$
        \STATE
        \STATE \textbf{Output:} Q-Load Sequence QSeq, K-MAC Sequence KSeq

        \STATE QSeq = [], KSeq = [], Head-id $i_h=0$, state = \textit{intoHD}  \myCOMMENT{Init}
        
        \WHILE{True}
        \IF{state is \textit{intoHD}}
            \IF{$HT_i$ is HEAD}
                \STATE QSeq.Update( $\text{Q}_{\text{i,HEAD}} + \text{Q}_\text{i,GLOB}$ )
            \ELSIF{$HT_i$ is TAIL}
                \STATE QSeq.Update( $\text{Q}_{\text{i,TAIL}} + \text{Q}_{\text{i,GLOB}}$ )
            \ENDIF
            \STATE \myCOMMENT{Finish reading K of head $i_{h-1}$}
            \STATE KSeq.Update( \texttt{Kid[$i_{h}-1$, $N-S_h$:]} ) 
            \STATE state $\gets$ midstHD
        \ELSIF{state is \textit{midstHD}}
            \STATE KSeq.Update( \texttt{Kid[$i_h$, $S_h$:$N-S_h$]} )
            \STATE state $\gets$ outtaHD
            \STATE $i_h \gets i_h + 1$
        \ELSIF{state is \textit{outtaHD}}
            \IF{$HT_i$ is HEAD}
                \STATE QSeq.Update( $\text{Q}_{\text{i,TAIL}}$ )
            \ELSIF{$HT_i$ is TAIL}
                \STATE QSeq.Update( $\text{Q}_{\text{i,HEAD}}$ )
            \ENDIF
            \STATE KSeq.Update( \texttt{Kid[$i_h$, $:S_h$]} )
            \STATE state $\gets$ \textit{intoHD}
        \ENDIF

        \IF{$i_h == N_h$}
            \STATE break
        \ENDIF

        \ENDWHILE
        
        \RETURN QSeq, KSeq
        
    \end{algorithmic}
}
\end{algorithm}

For selective token mechanism, a straightforward approach to reduce energy is to gate compute units when loading operands (e.g.,trivial Keys). 
Yet, such pruning brings marginal benefits because it fails to fully exploit system utilization.
SATA goes beyond pruning to sparsity-aware inter-head scheduling, with the central idea of dynamically orchestrating Q-K operations across heads so that system modules (e.g., Systolic array PEs or CIM tiles) remain fully utilized. 

Unlike prior works that fix Ks as the stationary operand~\cite{yang2020retransformer, fu2023p, tu2022trancim, tu2023multcim}, SATA designates Qs to be stationary due to their low variance of arithmetic intensity. 
Specifically, each Q would always have the same amount of K access, while Ks behave otherwise~\cite{xiao2024ttst, zhao2019explicit}. 

The scheduling flow can be formalized as a Finite State Machine~(FSM). Each state corresponds to a different situation scheduling HEAD/TAIL/GLOB sets: 
\begin{itemize}
    \item init: Load major Qs, which mean HEAD\&GLOB for condition \underline{HEAD}, or TAIL\&GLOB for condition \underline{TAIL}.
    \item intoHD: Launch MatMul with the first $[0:S_h-1]$ sorted Ks while loading minor Qs (TAIL Qs for condition \underline{HEAD}, or HEAD Qs for condition \underline{TAIL}). 
    This is possible since sorting has granted HEAD~(TAIL) Qs would not require the last~(first) $S_h$ sorted Ks to generate score matrix, which enabled parallel execution. 

    \item midstHD: This state is activated only when $S_h$ is not half the tile size. Ks with sorted indices $[S_h:N-S_h]$ perform MAC with every Q to preserve model accuracy.

    \item outtaHD: Conclude MAC with Ks indexed $[N-S_h:N]$ and start loading major Qs for the next head. 
    Since the current major Qs need not access the leftover Ks, they can be safely retired and release storage capacity.
    \item wrapGLOB: For heads in \underline{GLOB} state, perform Q-K MatMul in a 'Load Q then MAC with K' manner.
\end{itemize}



The scheduling stops when all LOCAL heads are consumed. For the remaining \underline{GLOB} heads, the scheduler reverts to conventional computation flow. 
In experimentation, most of the \underline{GLOB} heads can escape \underline{GLOB} state with a smaller $S_h$. The profiled data shows $<0.1\%$ \underline{GLOB} heads are observed among 2K Traces from TTST.  

Algo.~\ref{alg:schedule} summarizes scheduling. 
Note that edge transition and GLOB wrapping up are omitted for brevity. 
Sorting and scheduling are visualized in Fig.~\ref{fig:schedule_temporal}.



\subsection{Scale to Larger Sequence Length}
\label{sec:alg_longseq}

Scaling to longer sequences is one of the critical hurdles in both the algorithmic and hardware communities. A growing sequence length $N$ incurs quadratic growth in Q-K MatMul and extensive off-chip traffic.

SATA introduces tiling within each head to create smaller sub-blocks (e.g., $16\times16$ tiles). 
Each tile is executed like a 'subhead'. 
Sorting would be conducted across Q-folds while fold-wise Ks are reused. 
A similar process would then repeat across K-folds and complete the score matrix. This brings buffer demand to become manageable.

In the original sorting scheme, every Key and Query is determined indispensable. 
This is invalidated in tiled subheads. 
As such, zero-skip is introduced to the tiled selective masks. 
The scheduler identifies redundant Query(Key) in the tiled matrix by a column(row)-wise reduction AND operation. 
Together, tiling and zero-skip enable SATA to scale to long sequences  with preserved locality, while preventing the quadratic growth to overwhelm hardware.

\subsection{Hardware Implementation}
\label{sec:hw_describe}

SATA scheduling scheme is implemented with a lightweight scheduler integrated with compute engine. 
Its functionality is to coordinate memory traffic in and out of the compute unit.  

We decompose the functionalities of SATA and organize them into digital modules as in Fig.~\ref{fig:integrated_framework}a. 
Zero-unit detects and filters redundancy. 
Dot-product engine updates the similarity score and sorts Keys with a priority encoder.
Key order is staged in a FIFO during sorting.
Query order is written to another FIFO based on the classification result. 
FIFOs can be read by computational parts for operand retrieval.  

The most energy and runtime consuming step for scheduling is dot products~(line 7-12 in Algo.~\ref{alg:sort}). 
Instead of directly computing similarity and updating the dummy vector~(Eq.~\ref{eq:dis_raw}), SATA saves the cumulative distance between the \textit{Dummy} vector and all mask columns ($QK[:,i],i\in[0,N-1]$). 
Whenever a Key index $j$ is sorted, the Psum Registers are incremented by the binary dot-product result between $QK[:,j]$ and all unsorted columns~(Eq.~\ref{eq:dis_op}). 
\vspace{-3pt}
\begin{align}
    Distance_i &= Dummy^T \cdot QK[:,i] \label{eq:dis_raw} \\
    \text{Psum-Reg}[i] &{+}{=}\; QK[:,i]^T \cdot QK[:,j], \quad i \notin Kseq \label{eq:dis_op}
\end{align}

This essentially eliminates the repetitive MAC operation between sorted Ks and unsorted ones, and delivered design with better PPA metrics. 
The overhead of the scheduler is discussed in Sec.~\ref{sec:scheduler_overhead}.

\section{Evaluation}
\label{sec:eva}
This section evaluates SATA's efficiency, scalability, hardware overhead, and compatibility with SOTA accelerators.
We first introduce the estimation framework and workloads, then present throughput, energy, and SOTA-compatibility results.

\subsection{Estimation Framework and Testing Workloads}

\begin{figure*}[t!]
    \centering
    \includegraphics[width=.79\linewidth]{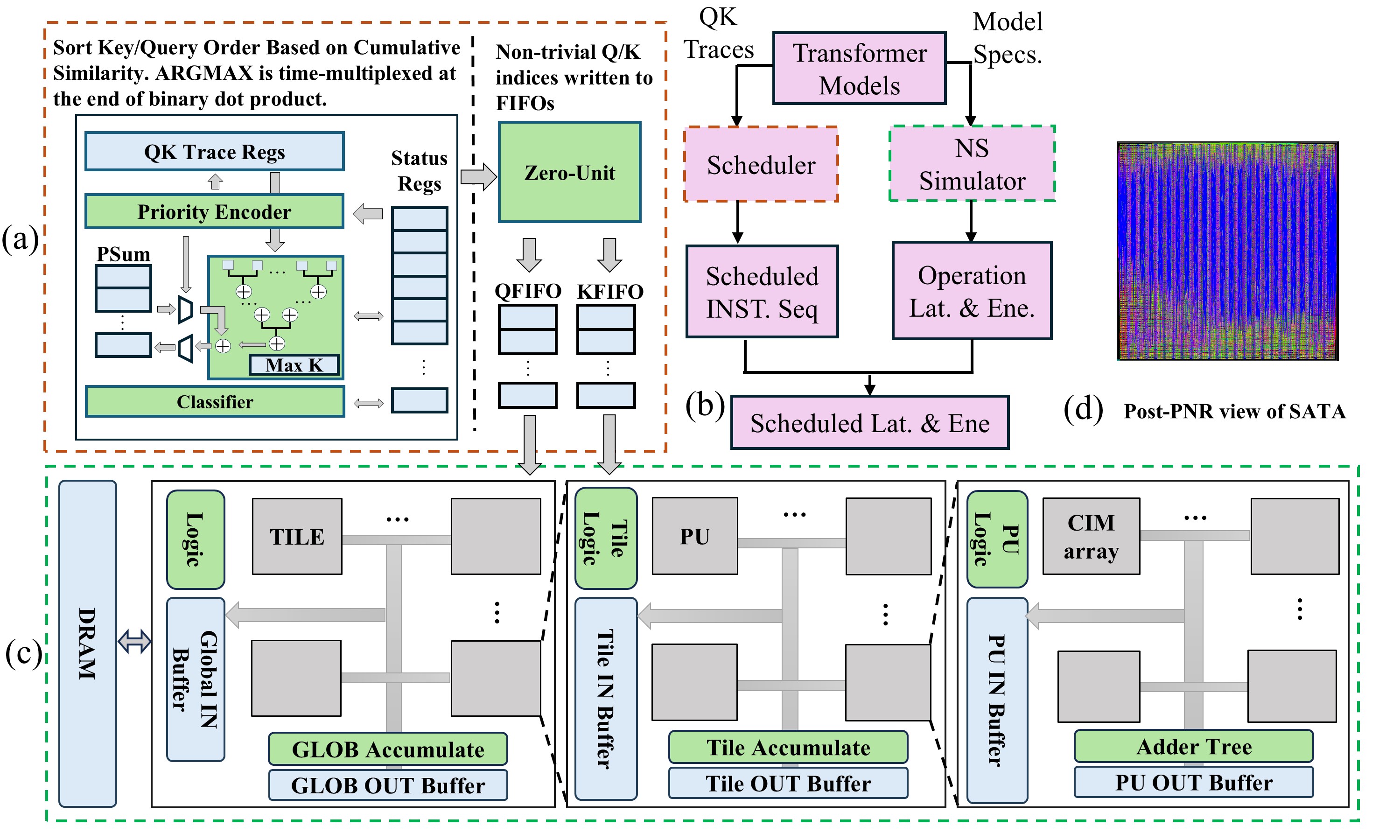}

    \caption{\textbf{SATA estimation framework.} 
    (a) Schematic view of the proposed scheduler. Status-Regs control the scheduling process and stage intermediate data. 
    (b) Workflow of SATA estimation framework. 
    (c) Homogeneous CIM-centric computational system initialized by NeuroSim. Input activations start from DRAM, execute MAC operation on sub-arrays, and are transferred back to DRAM to finish computation. 
    (d) Post-PNR figure of SATA.
    }
    \label{fig:integrated_framework}
\end{figure*}

\begin{table*}[htb]
\caption{Workload Specification \& Post-Schedule Statistics.}
\centering
\begin{tabular}{c|c|c|c|c|c|c|c|c}
\hline

\textbf{TopK Model} & \textbf{EMB-DIM ($D_k$)} & \textbf{K/\#Token} & \textbf{0-Skip} & \textbf{Dataset} & \textbf{GlobQ\%} & Tile Size($S_f$) & Avg Heavy-Size ($S_h$) & Avg \#($S_h$-=1)\\
\hline
TTST\cite{xiao2024ttst} & 65536 & 15/30 & 0 & 
~\cite{cheng2017remote} & 24.2\% & $N$ & 0.463$N$ & 1.55 \\
\hline
KVT-DeiT-Tiny~\cite{wang2022kvt} & 64 & 50/198 & 1 & 
~\cite{deng2009imagenet} & 33.3\% & 0.11$N$ & 0.053$N$ & 0.62 \\
\hline
KVT-DeiT-Base~\cite{wang2022kvt} & 64 & 64/198 & 1 & 
~\cite{deng2009imagenet} & 46.4\% & 0.11$N$ & 0.051$N$ & 1.38\\
\hline
DRSformer~\cite{chen2023learning} & 4800 & 12/48 & 1 & 
~\cite{yang2017deep} & 14.8\% & 0.125$N$ & 0.062$N$ & 0.05\\

\hline
\end{tabular}
\label{tab:model_table}
\end{table*}
\subsubsection{PPA estimation}
We access SATA with an estimation framework, which models both algorithmic behavior and hardware activities. 
Evaluated system consist of the computational CIM module~(Fig.~\ref{fig:integrated_framework}c) and the scheduler~(Fig.~\ref{fig:integrated_framework}a). 
We implement scheduler in SystemVerilog and synthesized it with TSMC65nm process in Design Compiler. 
Place and Route is done with IC Compiler2. Both SATA and CIM adopt 1GHz frequency clock rate.

CIM module is estimated with NeuroSim, which is validated and calibrated against post-silicon measurements~\cite{lu2021neurosim}. 
The CIM architecture is a multi-level, homogeneous system that takes both caches and interconnect into consideration (Fig.~\ref{fig:integrated_framework}c). 
We supplement the original dense CIM engine with SATA to quantify its throughput and energy benefit.

We size the CIM subarray to $32\times32$. 
The technology is adopting 65 nm process metadata. 
For additional system configuration details, we redirect interested readers to~\cite{NSUserManual}. 
We performed Design Space Exploration (DSE) on the SATA configuration to ensure optimal performance is delivered.

The estimation workflow~(Fig.~\ref{fig:integrated_framework}b) starts with trace extraction. 
Scheduler produces operator with QK traces.
The instructions (RD/MAC, WR/Load) are combined with a customized NeuroSim codebase to estimate latency and energy. 

\begin{figure*}[t]
    \centering
    \begin{subfigure}[t]{0.34\textwidth}
        \centering
        \includegraphics[width=\linewidth]{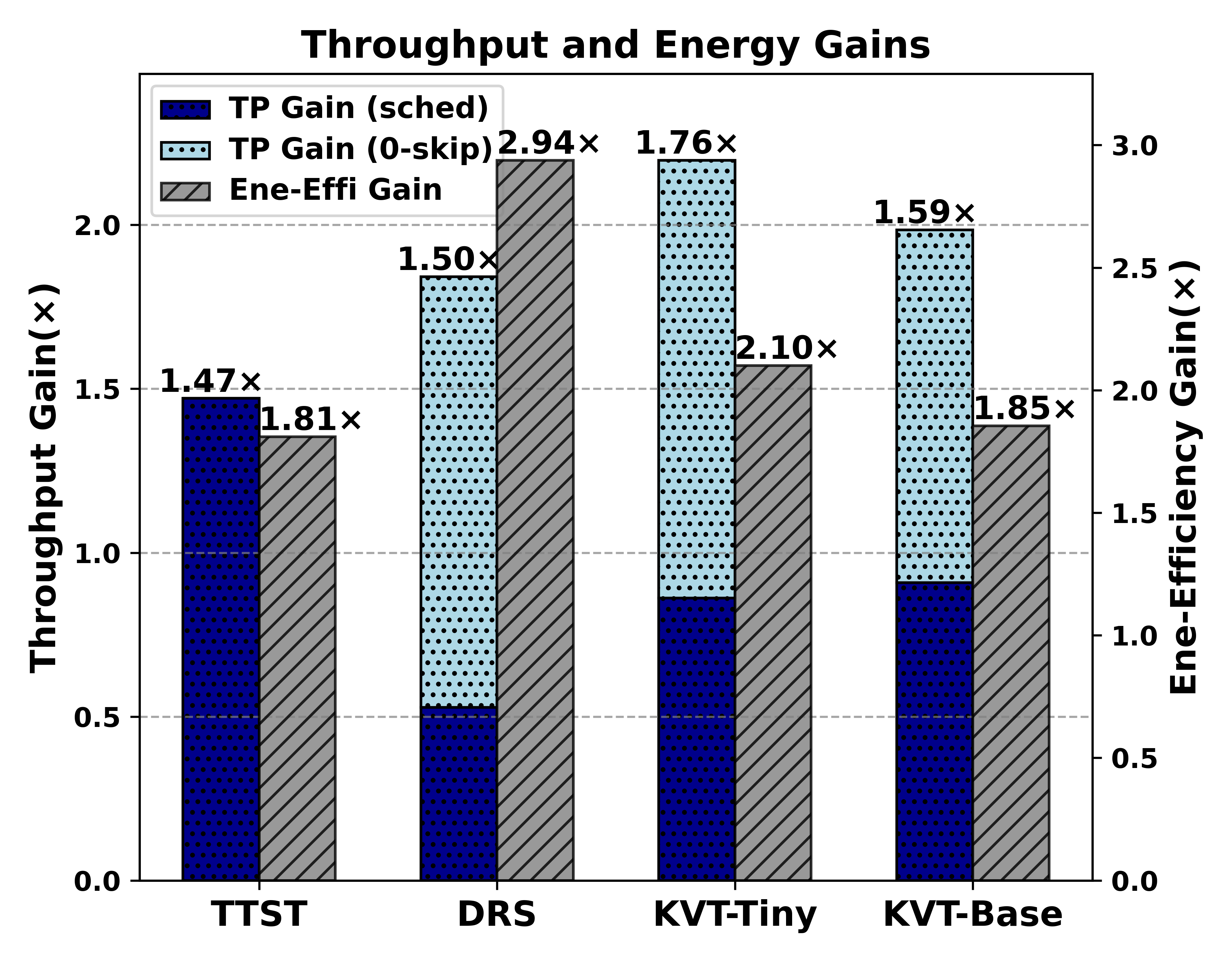}
        \caption{QK throughput and energy efficiency gain of SATA. The cost of QK index compute and scheduler has been incorporated.}
        \label{fig:QK_gain}
    \end{subfigure}
    \hfill
    \begin{subfigure}[t]{0.29\textwidth}
        \centering
        \includegraphics[width=\linewidth]{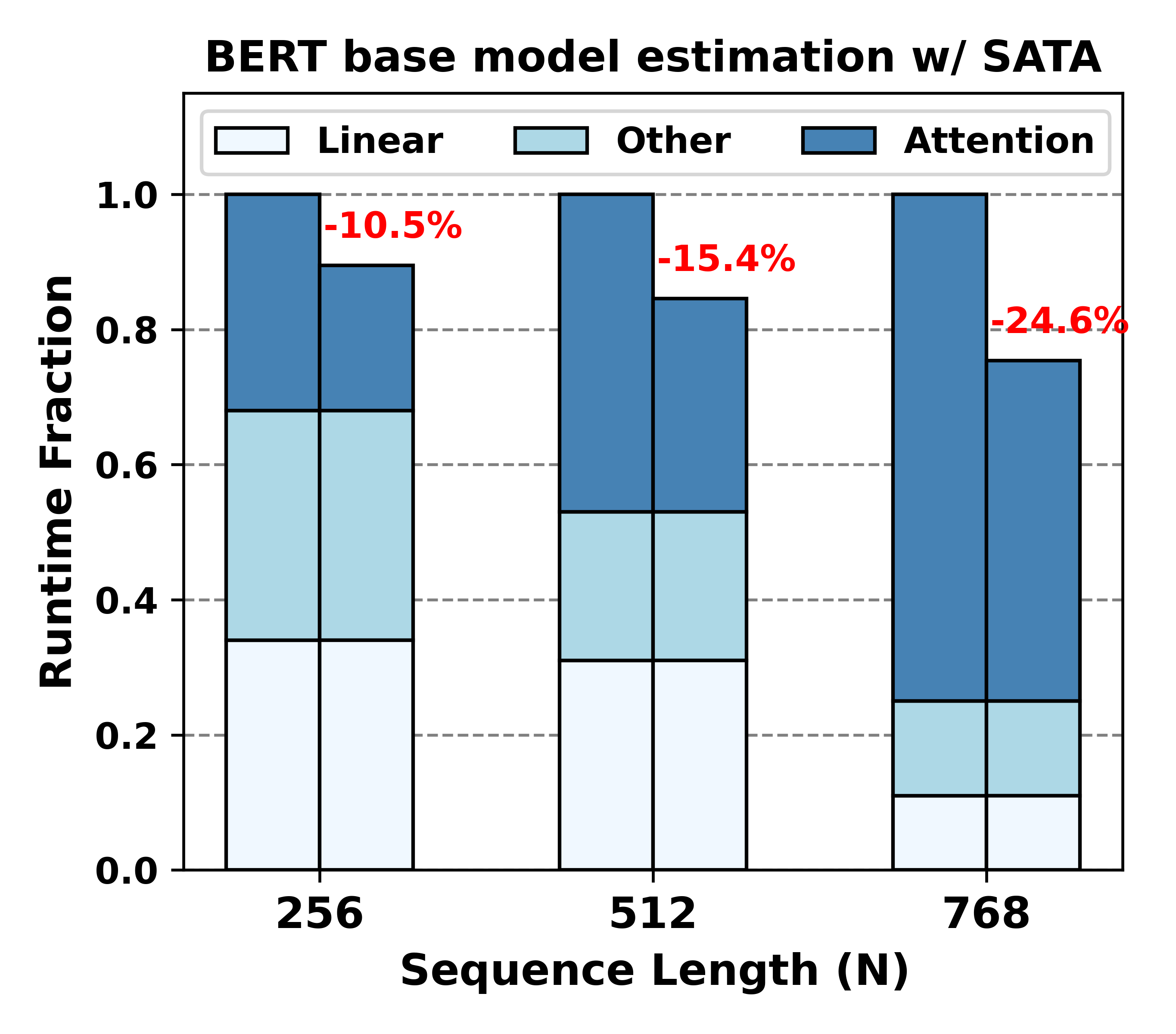}
        \caption{Normalized BERT based model estimation with SATA integration~\cite{zhou2022energon}. }
        \label{fig:selfattn_gain}
    \end{subfigure}
    \hfill
    \begin{subfigure}[t]{0.31\textwidth}
        \centering
        \includegraphics[width=\linewidth]{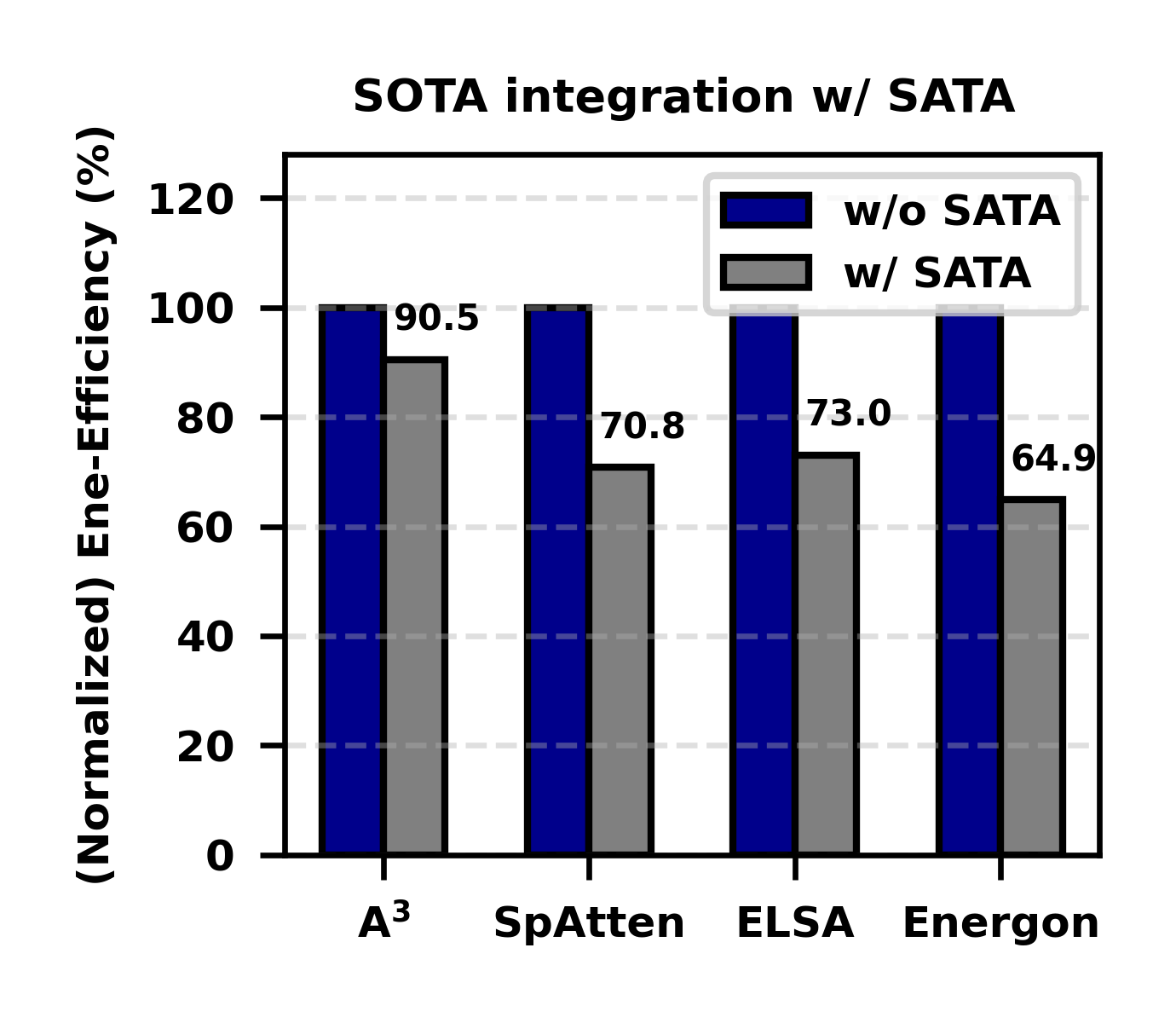}
        \caption{Energy-efficiency gain by integrating SATA into SOTA accelerators.}
        \label{fig:SOTA_intgt}
    \end{subfigure}

    \caption{\textbf{SATA evaluation results} for (a) QK throughput, energy efficiency gain for different workloads; (b) Self-Attention runtime reduction with SATA; and (c) SOTA energy-efficiency improvement with SATA.}
    \vspace{-3pt}
    \label{fig:sata_estimate}
\end{figure*}

\paragraph{Throughput}
The throughput benefit of SATA arises from reduced modules idleness after interleaving Q/K operations. 
We extract metadata of data transmission (DT) and Compute (Comp) latency from NeuroSim.
The latency($\tau$) for a scheduled time step that reads~(MACs) $x$ Ks and writes~(loads) $y$ Qs is estimated as
\vspace{-0.8em}
\begin{align}
\tau_i &= \min (\tau_{\text{RD,DT}} \cdot \textit{\textbf{x}}, \tau_{\text{WR,ARR}} \cdot \textit{\textbf{y}})  \notag \\
      &\quad + \min(\tau_{\text{RD,COMP}} \cdot \textit{\textbf{x}}, \tau_{\text{WR,DT}} \cdot \textit{\textbf{y}})
\end{align}
The latency of scheduling is negligible compared to compute~($<5\%$) and can be hidden through pipelining. 

\paragraph{Energy Estimation}
NeuroSim estimates energy assuming dense MatMul. 
In scheduled operations, only a subset of tiles perform MAC, while the remaining can update weights~(Qs). 
Assume a head is sorted to have $a$ HEAD Qs, $b$ TAIL Qs, and $c$ GLOB Qs ($S_h<\frac{N}{2}$). 
Then Keys indexed $[N-S_h:N]$ would bypass $a$ Qs; Keys indexed $[0:S_h]$ would bypass $b$ Qs. 
MACs are assumed to be conducted in a dense manner, albeit in a subset of tiles. 
The energy of SATA is extracted from EDA reports.


\subsubsection{Selective Transformer Models}
We benchmark SATA against k-NN Attention~\cite{wang2022kvt}, TTST~\cite{xiao2024ttst}, and DRSformer~\cite{chen2023learning}, all of which adopted TopK attention.
Model parameters are listed in Tab.~\ref{tab:model_table}. 
Values of $K$ are chosen, such that the model performance degradation is negligible. 

\subsection{Performance and Algorithmic Parameter Analysis}
\label{sec:gain_estimate}

Fig.~\ref{fig:QK_gain} shows the scheduled Q-K attention performance result for TTST. 
The throughput benefit arises from parallelized scheduled operation, hence improved hardware utilization, while the energy reduction comes from pruned MAC operations. 
We initialize $S_h = \frac{N}{2}$ and $\theta = \frac{N}{2}$.
$S_h= \frac{N}{2}$ represents the optimistic case, though relaxation often happens to escape \underline{GLOB} state. 
In practice, the average post-schedule $S_h$ stays close to the optimal condition~(50\% tile size), as reported in Tab.~\ref{tab:model_table}.
The average $S_h$ decrement times are also given in Tab.~\ref{tab:model_table}. 
This incurs minimal overhead, as the latency and energy consumption of classification is minor in scheduler. 

We also integrate the QK-index computation and SATA scheduling cost into estimation shown in Fig.~\ref{fig:QK_gain} ~\cite{wang2021spatten, zhou2022energon}. 
SATA achieves a peak throughput gain of 1.76$\times$ and an energy efficiency benefit of 2.94$\times$. 
Moreover, SATA is compatible with MatMul engine beyond CIM. 
A preliminary TTST test on a SATA-enhanced systolic array platform, modeled with ScaleSIM~\cite{raj2025scale}, demonstrated a 3.09$\times$ throughput gain, with stall cycles reduced from 90.4\% to 75.2\%.

\subsection{Scaling capability}

In SATA, long vectors are difficult to sort algorithmically, and the scheduler itself becomes prohibitive due to the logarithmic growth of tree structure.
We pick tile size ($S_f$) such that SATA delivers the highest performance. The $S_f$ values for different workloads are recorded in Tab.~\ref{tab:model_table}.

As $S_f$ decreases, throughput gain initially increases as SATA enables higher system utilization. 
When $S_f$ becomes even smaller, the benefit from zero-skip starts to dominate (exceeding $50\%$). 
Redundant operands will not be pushed into FIFOs, which means higher fraction of trivial operands inherently makes scheduling contributions less significant.

\subsection{Scheduler Overhead}
\label{sec:scheduler_overhead}


We evaluate the overhead of SATA scheduler compared to computational module. 
The cost is negligible with large embedding dimensions but
scales quadratically with tile size (register array) and logarithmically with tree-style modules. 
Scheduling latency can be effectively hidden through pipelining as long as scheduler takes less time than Q-K MatMul.


Generally, latency overhead remains minor~($<5\%$) when $D_k\geq64$, or when $S_f\leq24$, compared to an optimized CIM core. 
In terms of energy, a minor cost~($<5\%$) assumption fails when $D_k<32$ or $S_f>28$, which already falls out of the optimal gain setting reported in Tab.~\ref{tab:model_table}.  

\subsection{Integrating SATA to Existing Accelerators}


Previous accelerators~\cite{ham20203, ham2021elsa, wang2021spatten, zhou2022energon} execute sparse Q-K MAC after index acquisition. 
Though parallel hardware or pipelined designs improve efficiency, their sparsified nature remain underutilized. 
By integrating SATA, additional energy and throughput benefits can be achieved through localized operand access. 
Fig.~\ref{fig:SOTA_intgt} presents an energy efficiency estimation by enhancing SOTA designs with the proposed hardware. 
On average, SATA can provide 1.34$\times$ energy efficiency and 1.3$\times$ throughput gain after integration. 
$A^3$'s recursive search dominates runtime overhead and shows limited improvement.

\vspace{-0.02in}
\section{Conclusion}
\label{sec:conclude}
We presented SATA, a sparsity-aware scheduling framework that enhances selective-token attention by improving operand locality, reducing redundant memory access, and sustaining high utilization. With tiling and zero-skip extensions, SATA scales efficiently to long sequences. Evaluations with silicon-validated simulations and selective-attention traces demonstrate throughput and energy efficiency improvements with low overhead. By bridging algorithmic sparsity with hardware utilization, SATA provides a locality-centric, proactive scheduling that complements existing accelerators and paves the way for more scalable and efficient Transformer deployment.


\section*{Acknowledgements}
This work was supported in part by CoCoSys, one of seven centers in JUMP 2.0, a Semiconductor Research Corporation (SRC) program sponsored by DARPA.

\bibliographystyle{ieeetr}
\bibliography{ref}

\end{document}